\documentclass[aps,prl,twocolumn,showpacs,superscriptaddress,groupedaddress]{revtex4}  % for review and submission
\usepackage{graphicx}  % needed for figures
\usepackage{dcolumn}   % needed for some tables
\usepackage{bm}        % for math
\usepackage{amssymb}   % for math

\begin{document}

% The following information is for internal review, please remove them for submission
%\widetext
%\leftline{Version xx as of \today}
%\leftline{Primary authors: Joe E. Physics}
%\leftline{To be submitted to (PRL, PRD-RC, PRD, PLB; choose one.)}
%\leftline{Comment to {\tt d0-run2eb-nnn@fnal.gov} by xxx, yyy}
%\centerline{\em D\O\ INTERNAL DOCUMENT -- NOT FOR PUBLIC DISTRIBUTION}

% the following line is for submission, including submission to the arXiv!!
%\hspace{5.2in} \mbox{Fermilab-Pub-04/xxx-E}

\title{Theory of Skyrmionic Diffusion:

Hidden Diffusion Coefficients and Breathing Diffusion}

\author{E. Tamura}
\email[]{tamura@spin.mp.es.osaka-u.ac.jp}
\affiliation{Graduate School of Engineering Science, Osaka University, Toyonaka, Osaka 560-8531, Japan}

\author{Y. Suzuki}
\email[]{suzuki-y@mp.es.osaka-u.ac.jp}
\affiliation{Graduate School of Engineering Science, Osaka University, Toyonaka, Osaka 560-8531, Japan}
\affiliation{Center for Spintronics Research Network, Osaka University, Toyonaka, Osaka 560-8531, Japan}

\date{\today}

\begin{abstract}
Time evolution of the position-velocity correlation functions (PVCF) plays a key role
 in a new formalism of Brownian motion.
A system of differential equations, which governs PVCF, is derived for magnetic Skyrmions 
on a 2-dimensional magnetic thin film with thermal agitation.
In the formalism, a new type of diffusion coefficient is introduced
which does not come out in the usual diffusion equations.
The mean-square displacement (MSD) is obtained from the PVCF and found 
that it oscillates in time when the damping constant is small. 
It is also shown, even for a structureless particle, that the famous Ornstein-F\"urth formula should be corrected taking a proper initial value of PVCF into account. 
\end{abstract}

\maketitle

%\section{\label{sec:level1}First-level heading}
% sections are not used for PRL papers
Skyrmionic diffusion is viewed as a motion of the topologically protected local excitations in a thermally
fluctuating magnetic medium. When the form of the excitations changes marginally in a moderate thermal condition,
the excitations {\it Skyrmions} act as  small particles immersed in a liquid, and therefore, 
established methods for analyzing Brownian motion can be applied. 
The standard approach to the problems of free diffusion has been based on Einstein's classical result 
for the mean-square displacement (MSD) \cite{Einstein1956}, $ \lim_{t \to \infty} \langle x^2(t) \rangle = 2D_{xx}t $, and {\it ditto} for $y$.  For an isotropic 2-dimensional case the other relation, $ \langle x(t)y(t) \rangle = 0 $, holds, because of its rotational symmetry.
Differentiating these relations with respect to time $t$, the equations are expressed in a compact form,
$$ \begin{array}{lr} D_{xx}=D_{yy} & ; \quad D_{xy}=-D_{yx} \end{array} , \eqno(1) $$
where
$$D_{ij}\equiv \lim_{t \to \infty} \langle x_i(t) v_j(t) \rangle \,\,\, ; \quad (i=x,y) . \eqno(2) $$
The diffusion coefficients $D_{ij}$ are described by means of the position-velocity correlation functions (PVCF) 
 and  are the coefficients of 2nd-order
partial derivatives $ \partial^2/{\partial x_i \partial x_j} $ in the diffusion equation.  Although it is
obvious that the terms of $D_{xy}$ and $D_{yx}$ cancel each other out in the diffusion equation, 
it dose not mean at all that their values vanish. In this Letter, we develop a new formalism of the diffusion
process, specifically suitable for skyrmionic diffusion, where the diffusion coefficients $D_{xy}$ and $D_{yx}$ play an important role. Furthermore, we give a correction to the century old famous Ornstein-F\"urth formula 
 \cite{Ornstein1919,Langevin2017} which has been used for an expedient determination of the MSD.

The motion of Skyrmions is governed by the generalized Thiele-Langevin equation \cite{Thiele1973,Schutte2014,Milta2018}.  For the 2-dimensional velocity
$\mbox{\boldmath $v$}=v_x\hat{\mbox{\boldmath $x$}}+v_y\hat{\mbox{\boldmath $y$}}$, the equation is given in the
following form
$$ m\left( \begin{array}{cc} \cos\vartheta & -\sin\vartheta \\ \sin\vartheta & \cos\vartheta \end{array} \right)
\frac{d\mbox{\boldmath $v$}}{dt}=-\alpha D \mbox{\boldmath $v$}+\mbox{\boldmath $G$}\times\mbox{\boldmath $v$}
+\mbox{\boldmath $R$}(t), \eqno(3) $$
where $\alpha$ is the Gilbert damping constant, $D$ the dissipation dyadic,  $\mbox{\boldmath $G$}=-G\hat{\mbox{\boldmath $z$}}$ the gyromagnetic coupling and 
$\mbox{\boldmath $R$}(t)=R_x(t)\hat{\mbox{\boldmath $x$}}+R_y(t)\hat{\mbox{\boldmath $y$}}$ the random force 
due to thermal fluctuations. $\alpha D$ describes the friction and the stochastic property of $\mbox{\boldmath $R$}(t)$ has been obtained from the dynamics of a classical magnetic moment with thermal agitation \cite{Milta2018,sLLG1998} as
% $$ \left\{ {\begin{array}{rl} \langle \mbox{\boldmath $R$}(t) \rangle & = \mbox{\boldmath $0$} \\ 
% \langle R_i(t) R_j(t) \rangle & = C_{\rm R} \delta_{ij} \delta (t-t') \end{array} $$
$$ \begin{array}{lr} \langle \mbox{\boldmath $R$}(t) \rangle = \mbox{\boldmath $0$} & ; \quad 
 \langle R_i(t) R_j(t) \rangle = C_{\rm R} \delta_{ij} \delta (t-t') , \end{array} \eqno(4)$$
where the variance of the random force is given by $ C_{\rm R}=2k_{\rm B}T\alpha D$. 
It is worthy to note that $C_{\rm R}$ is defined {\it a priori} and not determined {\it a posteriori} in the
theory of skyrmionic diffusion.
The Thiele equation was originally formulated for the steady motion of magnetic domains \cite{Thiele1973},
that is, the left-hand side of the Eq.(3) does not appear. It is a natural generalization to add a term of first
derivative of the velocity that is not constant anymore. The coefficient in the term is presumably called 
mass that is not necessarily a scalar but tensor. Since our system under consideration is 2-dimensionally 
isotropic, the tensor should have a form employed in the Eq.(3).

Since $\mbox{\boldmath $R$}(t)$ is isotropic, that is, independent of the rotations around 
$\hat{\mbox{\boldmath $z$}}$-axis, the Eq.(3) can be reduced to the form which does not explicitly depend
on $\vartheta$
$$\frac{d\mbox{\boldmath $v$}}{dt}=-\gamma \mbox{\boldmath $v$}-\Gamma \hat{\mbox{\boldmath $z$}}\times\mbox{\boldmath $v$}
+\frac{\mbox{\boldmath $R$}(t)}{m} , \eqno(5) $$
where $\gamma$ and $\Gamma$ are defined by
 $$ \left\{ \begin{array}{c} \gamma \equiv \mbox{\Large{$\frac{\alpha D}{m}$}} \cos\vartheta 
 + \mbox{\Large{$\frac{G}{m}$}} \sin\vartheta , \\ \noalign{\vskip0.2cm}
 \Gamma \equiv \mbox{\Large{$\frac{G}{m}$}} \cos\vartheta -\mbox{\Large{$\frac{\alpha D}{m}$}} \sin\vartheta \end{array} . \right. \eqno(6) $$
 Introducing a standard Wiener process $\{\mbox{\boldmath $W$}(t),t \geq 0 \}$ and its increment 
 $$  \sqrt{C_{\rm R}}d\mbox{\boldmath $W$}(t) = \mbox{\boldmath $R$}(t)dt , $$
 we rewrite the Eq.(5) in terms of stochastic differential equation (SDE) \cite{Langevin2017}. 
 In SDE terminology, It\^o's form of Eq.(5) together with the position increment expressed by 
 the velocity and a time increment, is given by
 $$ \left\{ {\begin{array}{l} d\mbox{\boldmath $v$}= \left(-\gamma \mbox{\boldmath $v$}
-\Gamma \hat{\mbox{\boldmath $z$}}\times\mbox{\boldmath $v$}\right)dt  
 + \mbox{\Large{$\frac{\sqrt{C_{\rm R}}}{m}$}}\, d\mbox{\boldmath $W$} , \\ \noalign{\vskip0.2cm}
 d\mbox{\boldmath $x$}= \mbox{\boldmath $v$}dt , \end{array}} \right. \eqno(7) $$ 
 with
 $$ \begin{array}{lr} \langle d\mbox{\boldmath $W$} \rangle = \mbox{\boldmath $0$} & ; \quad 
 \langle dW_i dW_j \rangle =  \delta_{ij} dt . \end{array} \eqno(8) $$
 It is important to note that a different stochastic integration, such as Stratonovich's form, does not alter 
 the results since the coefficient of  $d\mbox{\boldmath $W$}$ is a constant.  We proceed further to manipulating
 equations by applying the product rule of It\^o's stochastic calculus. It is a rule for the increment of a product consisted of 2 stochastic variables, $d(fg)=(df)g+fdg+dfdg$, and the last term, the product of 2 increments, is not negligible owing to 
the equality Eq.(8). 
%$$d\left(\mbox{\boldmath $v$}^2 \right) = 2\mbox{\boldmath $v$}\cdot d\mbox{\boldmath $v$} + 
%          d\mbox{\boldmath $v$}\cdot d\mbox{\boldmath $v$}$$ 
Accordingly, we construct the following 3 SDEs from the Eq.(7).
\begin{widetext}
$$ \left\{ {\begin{array}{rl} d(\mbox{\boldmath $x$}\cdot\mbox{\boldmath $v$}) &=\left( -\gamma \mbox{\boldmath $x$}\cdot\mbox{\boldmath $v$}
+\Gamma \hat{\mbox{\boldmath $z$}}\cdot(\mbox{\boldmath $x$}\times\mbox{\boldmath $v$})+\mbox{\boldmath $v$}^2 \right)dt  
 + \mbox{\Large{$\frac{\sqrt{C_{\rm R}}}{m}$}}\,\mbox{\boldmath $x$}\cdot d\mbox{\boldmath $W$}, \\ \noalign{\vskip0.2cm} 
 d\left(\hat{\mbox{\boldmath $z$}}\cdot(\mbox{\boldmath $x$}\times\mbox{\boldmath $v$})\right)
 &= \left(-\gamma\hat{\mbox{\boldmath $z$}}\cdot(\mbox{\boldmath $x$}\times\mbox{\boldmath $v$})-\Gamma\mbox{\boldmath $x$}
\cdot\mbox{\boldmath $v$}\right)dt+\mbox{\Large{$\frac{\sqrt{C_{\rm R}}}{m}$}}\,\hat{\mbox{\boldmath $z$}}\cdot(\mbox{\boldmath $x$}
\times d\mbox{\boldmath $W$}) , \\ \noalign{\vskip0.2cm}
 d\left(\mbox{\Large{$\frac{1}{2}$}}\mbox{\boldmath $v$}^2 \right) &=\left( -\gamma \mbox{\boldmath $v$}^2
+\mbox{\Large{$\frac{C_{\rm R}}{m^2}$}} \right)dt+ \mbox{\Large{$\frac{\sqrt{C_{\rm R}}}{m}$}}\,\mbox{\boldmath $v$}\cdot d\mbox{\boldmath $W$}. 
 \end{array}} \right. \eqno(9) $$  
\end{widetext}
Then we take an average over all trajectories of $\mbox{\boldmath $W$}(t)$ and symbolize it by
$\langle \cdots \rangle $. 
 Since,  by definition, $ \langle\mbox{\boldmath $x$}\cdot d\mbox{\boldmath $W$}\rangle 
 = \langle\hat{\mbox{\boldmath $z$}}\cdot(\mbox{\boldmath $x$}\times d\mbox{\boldmath $W$})\rangle
 = \langle\mbox{\boldmath $v$}\cdot d\mbox{\boldmath $W$}\rangle =0 $  in accordance with the rule of 
It\^o's stochastic integration, we obtain a system of ordinary differential equations (ODE),
$$ \left\{ \begin{array}{rl} \mbox{\Large{$\frac{dQ_{\rm P}}{dt}$}} &=-\gamma Q_{\rm P}+\Gamma Q_{\rm S}+2Q_{\rm V}, \\
   \noalign{\vskip0.2cm}
                              \mbox{\Large{$\frac{dQ_{\rm S}}{dt}$}} &=-\Gamma Q_{\rm P}-\gamma Q_{\rm S}, \\
   \noalign{\vskip0.2cm}
                              \mbox{\Large{$\frac{dQ_{\rm V}}{dt}$}} &=-2\gamma Q_{\rm V}+ \mbox{\Large{$\frac{C_{\rm R}}{m^2}$}}, \end{array} \right. \eqno(10) $$
 for the 3 variables $Q_{\rm P}$, $Q_{\rm S}$ and $Q_{\rm V}$ which are defined by
 $Q_{\rm P}\equiv\langle\mbox{\boldmath $x$}\cdot \mbox{\boldmath $v$}\rangle$,
 $Q_{\rm S}\equiv\langle\hat{\mbox{\boldmath $z$}}\cdot(\mbox{\boldmath $x$}\times \mbox{\boldmath $v$})\rangle$ and 
 $Q_{\rm V}\equiv\langle\mbox{\boldmath $v$}^2\rangle /2$. $Q_{\rm P}$ ($Q_{\rm S}$) is the PVCF between
 parallel (perpendicular: {\it senkrecht}) position-velocity components. Recalling the Eqs.(1) and (2), they can 
 be  written more concretely, $ Q_{\rm P}(t) = \langle x(t)v_x(t) \rangle + \langle y(t)v_y(t) \rangle =2 \langle x(t)v_x(t) \rangle $ and $ Q_{\rm S}(t) = \langle x(t)v_y(t) \rangle - \langle y(t)v_x(t) \rangle =2 \langle x(t)v_y(t) \rangle $. Only $ Q_{\rm V}$ is coupled directly with  $C_{\rm R}$ and trading thermal energies with the random force, while $ Q_{\rm S}$ is not coupled directly with $ Q_{\rm V}$ and, therefore, describes dissipationless processes.

For the stationary or equivalently setting $t=\infty$ for the solutions of the Eq.(10), we have the diffusion coefficients 
$$ \begin{array}{rl}
   \left( \begin{array}{c} Q_{\rm P}(\infty) \\  Q_{\rm S}(\infty)\end{array}\right) &=
   \left( \begin{array}{c} D_{xx}+D_{yy} \\  D_{xy}-D_{yx} \end{array}\right)=\left( \begin{array}{c} 2D_{xx} \\  2D_{xy} \end{array}\right) \\
      \noalign{\vskip0.2cm}
    &=\mbox{\Large{$\frac{C_{\rm R}}{m^2}$}}\mbox{\Large{$\frac{1}{\gamma^2+\Gamma^2}$}} 
 %\left( \begin{array}{c} 1 \\ -\Gamma /\gamma \end{array}\right)  
  \left( \begin{array}{c} 1 \\ -\mbox{\Large{$\frac{\Gamma}{\gamma}$}} \end{array}\right),
   \end{array}  \eqno(11) $$ 
in addition to $$ Q_{\rm V}(\infty)=\frac{C_{\rm R}}{2\gamma m^2} . \eqno(12)$$
The former relationship given by the Eq.(11) is a Skyrmion version of Einstein's formula and the latter
given by the Eq.(12) is corresponding to Nyquist's theorem \cite{Langevin2017}. These 2 kinds of fluctuation-dissipation theorems
are derived simultaneously in our formalism. And it should be emphasized that the anomalous 
diffusion coefficients $ D_{xy} = -D_{yx} $ are present and have a sizable value, which we inferred earlier from general symmetry requirements and calculations of the velocity autocorrelation functions.
Hereafter we call them {\it gyro} diffusion coefficients. 
All the diffusion coefficients in the Eq.(11) are independent of the mass $m$ introduced in the Eq.(3),
while all the PVCF depend on it. Especially, the diffusion coefficients $ D_{xx} = D_{yy} $ do
not depend on the tensor component of mass $\vartheta$, since $m^2(\gamma^2+\Gamma^2)=(\alpha D)^2 + G^2 $
for any $\vartheta$.
When $\vartheta =0$ for the case of scalar mass,  the expression for $ D_{xx} (= D_{yy}) $ has been given 
\cite{Schutte2014,Milta2018} but for $ D_{xy} (= -D_{yx}) $ is shown for the first time,
  $$\left( \begin{array}{c} D_{xx} \\  D_{xy} \end{array}\right)
  =\mbox{\Large{$\frac{k_{\rm B}T}{\alpha D}$}} \frac{1}{1+\left( \mbox{\Large $\frac{G}{\alpha D}$} \right)^2}
  \left( \begin{array}{c} 1 \\ -\mbox{\Large{$\frac{G}{\alpha D}$}} \end{array}\right) . \eqno(13) $$
 An unusual form of the diffusion coefficients $ D_{xx} (= D_{yy}) $ can be explained by interaction with the gyro diffusion in our formalism. The ratio $G/\alpha D$ ($\alpha D/G$) varies possibly  $ 0.01 \sim 100 $ 
 ($ 100 \sim 0.01 $)  , while the character
of diffusion changes correspondingly from the normal to the suppressed \cite{Schutte2014,Milta2018}.  
The gyro diffusion coefficient does not appear in the diffusion equations. It means that the diffusion equations describe only processes with a dissipation. 
 
Wrapping up discussion on the stationary solutions,  we study now, the time evolution of the PVCF governed by the
Eq.(10).  Since the system is assumed to be in equilibrium and the velocity distribution stationary,
$Q_{\rm V}(t)=C_{\rm R}/2\gamma m^2 $,
the equations are reduced to a coupled non-homogeneous ODE,
$$ \frac{d}{dt} \left( \begin{array}{c} Q_{\rm P} \\  Q_{\rm S} \end{array}\right) =
  -\left( \begin{array}{cc} \gamma & -\Gamma \\ \Gamma & \gamma \end{array} \right) \left( \begin{array}{c} Q_{\rm P} \\  Q_{\rm S} \end{array}\right) + \left( \begin{array}{c} \mbox{\Large{$\frac{C_{\rm R}}{\gamma m^2}$}} \\ 0 \end{array}\right) . \eqno(14) $$
 
The general solutions of the equation are given in the form 
$$  \begin{array}{rl} 
\left( \begin{array}{c} Q_{\rm P}(t) \\  Q_{\rm S}(t) \end{array}\right) &=e^{-\gamma t}
\left( \begin{array}{cc} \cos{\Gamma t} & \sin{\Gamma t} \\ -\sin{\Gamma t} & \cos{\Gamma t} \end{array} \right) \times \\
   \noalign{\vskip0.2cm}
 & \qquad \quad  \times \left( \begin{array}{c} Q_{\rm P}(0^+) \\  Q_{\rm S}(0^+) \end{array}\right) 
  + \left( \begin{array}{c} q_{\rm P}(t) \\  q_{\rm S}(t) \end{array}\right) ,  \end{array} \eqno(15) $$
that is, solutions of the homogeneous equation with initial conditions $(Q_{\rm P}(0^+), Q_{\rm S}(0^+))$ 
and a particular solution
%$$ \left( \begin{array}{c} q_{\rm P}(t) \\  q_{\rm S}(t) \end{array}\right) = (D_{xx} + D_{yy}) 
%   \left[ \left( \begin{array}{c} 1 \\ -\mbox{\Large{$\frac{\Gamma}{\gamma}$}} \end{array}\right) + 
%   e^{-\gamma t}  \left( \begin{array}{c} -\cos{\Gamma t}+\mbox{\Large{$\frac{\Gamma}{\gamma}$}}\sin{\Gamma t} \\
%     \sin{\Gamma t}+\mbox{\Large{$\frac{\Gamma}{\gamma}$}} \cos{\Gamma t} \end{array}\right)
%   \right] . $$
%  
% $$ \begin{array}{rl} \left( \begin{array}{c} q_{\rm P}(t) \\  q_{\rm S}(t) \end{array}\right) & = (D_{xx} + %D_{yy}) 
%   \Biggr[ \left( \begin{array}{c} 1 \\ -\mbox{\Large{$\frac{\Gamma}{\gamma}$}} \end{array}\right) + \\
%      \noalign{\vskip0.2cm} & \qquad + 
%   e^{-\gamma t}  \left( \begin{array}{c} -\cos{\Gamma t}+\mbox{\Large{$\frac{\Gamma}{\gamma}$}}\sin{\Gamma t} \\
%    \sin{\Gamma t}+\mbox{\Large{$\frac{\Gamma}{\gamma}$}} \cos{\Gamma t} \end{array}\right)
%   \Biggr] , \end{array} \eqno(16)  $$ 
satisfying $(q_{\rm P}(0), q_{\rm S}(0))=0$.

Caution is demanded when we estimate the initial conditions, especially on $Q_{\rm P}(0^+)$ that in not
continuous at $t=0$. By definition,
$ Q_{\rm P}(0) \equiv \langle \mbox{\boldmath $x$}(0) \cdot \mbox{\boldmath $v$}(0) \rangle = 0$ 
because of $ \mbox{\boldmath $x$}(0) \equiv \mbox{\boldmath $0$}$. On the other hand, 
 at the infinitesimal positive time $t=0^+$
$$  \begin{array}{rl} 
  Q_{\rm P}(0^+) & = {\displaystyle \lim_{\delta t \to 0}}\langle \mbox{\boldmath $x$}(\delta t) \cdot \mbox{\boldmath $v$}(\delta t) \rangle \\
        \noalign{\vskip0.2cm} & 
   = {\displaystyle \lim_{\delta t \to 0}} \langle \mbox{\boldmath $x$}(\delta t) 
     \cdot {\displaystyle \frac{\mbox{\boldmath $x$}(\delta t)}{\delta t}} \rangle \\ 
        \noalign{\vskip0.2cm} & 
   = {\displaystyle \lim_{\delta t \to 0}} \langle
      {\displaystyle \frac{ \delta \mbox{\boldmath $x$}^2 }{\delta t}} \rangle
   = 2(D_{xx} + D_{yy}) \end{array}   \eqno(16) $$ 
where $ \delta \mbox{\boldmath $x$} = \mbox{\boldmath $x$}(\delta t) - \mbox{\boldmath $x$}(0) $
and the above equation is another definition of the normal diffusion coefficient \cite{Langevin2017}.
  And the other initial value $ Q_{\rm S}(0^+) =Q_{\rm S}(0) = 0 $, since $\langle x(t) y(t) \rangle =0 $
 for any $ t $.
   
The solution employing the above mentioned initial conditions, is given by
%$$ \left( \begin{array}{c} Q_{\rm P}(t) \\  Q_{\rm S}(t) \end{array}\right) = (D_{xx} + D_{yy}) 
%   \left[ \left( \begin{array}{c} 1 \\ -\mbox{\Large{$\frac{\Gamma}{\gamma}$}} \end{array}\right) + 
%   e^{-\gamma t}  \left( \begin{array}{c} \cos{\Gamma t}+\mbox{\Large{$\frac{\Gamma}{\gamma}$}}\sin{\Gamma t} \\
%     -\sin{\Gamma t}+\mbox{\Large{$\frac{\Gamma}{\gamma}$}} \cos{\Gamma t} \end{array}\right)
%   \right] . $$ 
 $$ \begin{array}{rl} \left( \begin{array}{c} Q_{\rm P}(t) \\  Q_{\rm S}(t) \end{array}\right) & 
 = (D_{xx} +   D_{yy}) 
   \Biggr[ \left( \begin{array}{c} 1 \\ -\mbox{\Large{$\frac{\Gamma}{\gamma}$}} \end{array}\right) \\
      \noalign{\vskip0.2cm} & \quad + 
   e^{-\gamma t}  \left( \begin{array}{c} \cos{\Gamma t}+\mbox{\Large{$\frac{\Gamma}{\gamma}$}}\sin{\Gamma t} \\
     -\sin{\Gamma t}+\mbox{\Large{$\frac{\Gamma}{\gamma}$}} \cos{\Gamma t} \end{array}\right)
   \Biggr] . \end{array} \eqno(17) $$
We draw the time evolution of the PVCF $(Q_{\rm P}(t), Q_{\rm S}(t))$  given by Eq.(17) in the Fig.1 and show 
 their $\alpha$ dependence. The curves are normalized by the value of $Q_{\rm P}(0^+)$ for $(\alpha D/G)=1$ under
which condition the diffusion coefficients $ D_{xx} (= D_{yy}) $ are maximal \cite{Milta2018} while the gyro diffusion coefficients $ D_{xy} (= -D_{yx}) $ take a maximum absolute value at $\alpha=0$. The features of the
PVCF differ drastically between the panels (a) and (b) across the condition $(\alpha D/G)=1$, although they
all converge to the diffusion coefficients when $t \to \infty $.

\begin{figure}[h]
\includegraphics[scale=0.4]{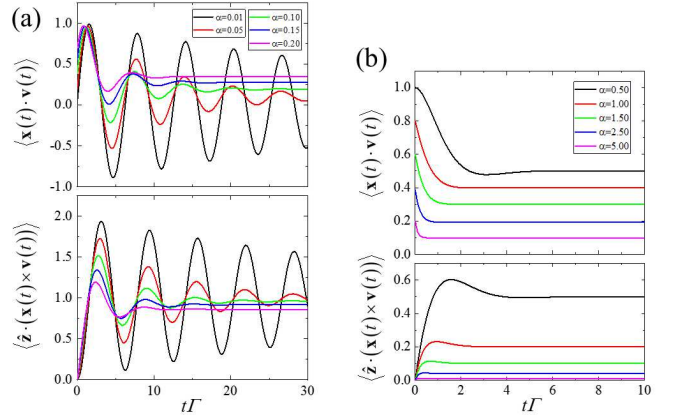}
\caption{\label{fig1} Position-velocity correlation function (PVCF) for the case $(D/G)=2$ and $\vartheta =0$.
  Panels (a) and (b) show the time evolution of the PVCF for $\alpha =0.01 \sim 0.2$ and $\alpha =0.5 \sim 5.0$ 
  respectively, and the upper and lower panels for $Q_{\rm P}$ and $Q_{\rm S}$. The sign of 
  $Q_{\rm S}$ is changed to fit into the figures.}
\end{figure} 

We are finally back to discussion on the MSD. Since we have 
$ d(\mbox{\boldmath $x$}^2)=2\mbox{\boldmath $x$} \cdot d\mbox{\boldmath $x$}
  =2\mbox{\boldmath $x$} \cdot \mbox{\boldmath $v$} dt $ from the Eq.(7), the MSD can be calculated by 
integrating $Q_{\rm P}(t)$ given in the Eq.(17) over the time period $t$,
$$ \begin{array}{rl}
 {\displaystyle \int_0^t\!\!} Q_{\rm P}(\tau)d\tau &= {\displaystyle \int_0^{\mbox{\boldmath $x$}(t)}\!\!\!\!}
 \langle\mbox{\boldmath $x$} \cdot d\mbox{\boldmath $x$}\rangle = \frac{1}{2}\langle\mbox{\boldmath $x$}^2(t)\rangle \\ 
 \noalign{\vskip0.2cm}
 & = 2D_{xx} \left[ t+{\displaystyle \frac{1}{\gamma}} ( 1-e^{-\gamma t}\cos{\Gamma t} ) \right] .  
 \end{array} \eqno(18)
$$
As is expected, the leading term is proportional to $t$ whose coefficient is the diffusion coefficient. 
In the Eq.(18) we use explicitly the 2-dimensional isotropy $ D_{xx} = D_{yy} $ , and hence factor 2
stands for the dimension. It should be noted that the tangential line does not pass through the origin
 and its ordinate intercept is positive and given by $2D_{xx}/\gamma$.
 
 In Fig. 2 and 3, we show the MSD given by Eq.(18) which are obtained from integrating the PVCF plotted
 in the panels (a) and (b) in the Fig. 1, respectively. 
 In the range of $\alpha = 0.01 \sim 0.2$, the diffusion coefficients $ D_{xx} (= D_{yy}) $
 are suppressed and become smaller for the smaller friction (Fig. 2). 
 The MSD for small $\alpha \le 0.1$ oscillates in time and repeats dilation and contraction 
 like a breathing which is, however, easily suppressed by a small tensor component of the mass
  $\vartheta =1$ deg. (see the black dashed line.)
 The breathing that we discuss, is 
 neither the Skyrmion-form breathing nor the breathing of Skyrmion distribution as a whole. 
 Nevertheless, we observe the breathing in the MSD on an average if we trace Skyrmions one by one. 
 
\begin{figure}[h]
\includegraphics[scale=0.4]{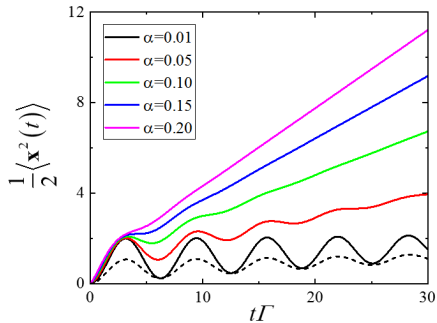}
\caption{\label{fig2a} Mean square displacement (MSD) for $\alpha =0.01 \sim 0.2$. The black dashed line shows
the MSD with the tensor component $\vartheta =1$ deg for $\alpha =0.01$.}
\end{figure}

 In the range of $\alpha \ge 0.5 $, the diffusion coefficients become smaller for 
 the larger friction (Fig. 3). The MSD curves shows a typical diffusion properties 
 described by Einstein's formula. From the ordinate intercept, we estimate the persistence time $1/\gamma$.
 (see the black dotted line.)
 
\begin{figure}[h]
\includegraphics[scale=0.4]{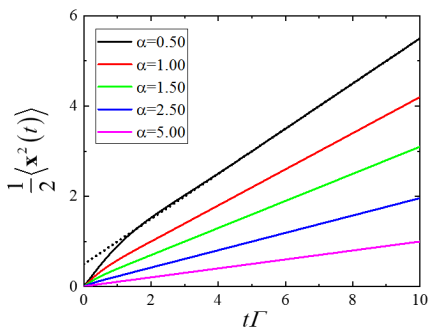}
\caption{\label{fig2b} Mean square displacement (MSD) for $\alpha =0.5 \sim 5.0$. The black dotted line shows
the tangential line of the MSD for $\alpha =0.5$.}
\end{figure}

 We calculate another physical quantity by integrating $Q_{\rm S}(t)$ which is presumably called
 the {\it mean rotation area}. 
$$ \begin{array}{rl} 
 {\displaystyle \int_0^t\!\!} Q_{\rm S}(\tau)d\tau &= {\displaystyle \int_0^{\mbox{\boldmath $x$}(t)}\!\!\!\!}
 \langle \hat{\mbox{\boldmath $z$}}\cdot (\mbox{\boldmath $x$} \times d\mbox{\boldmath $x$}) \rangle \\ 
 \noalign{\vskip0.2cm}
 & = 2D_{xy} \left[ t-{\displaystyle \frac{1}{\Gamma}} e^{-\gamma t}\sin{\Gamma t} \right] . \end{array} \eqno(19)
$$ The integral depends on the path of diffusing Skyrmions unlike the case of the MSD in the Eq.(18). 
  Hence the mean rotation area is well-defined only in the continuous limit. In the experiment,
  a time step of observations is chosen enough small to make the integral converge.  Otherwise,
  as is so in the Eq.(18), the gyro diffusion coefficient $ D_{xy} (= -D_{yx}) $ is  estimated by
   the values of the mean rotation area.

At the end, we compare our theory to the existing theory, the formulation for the MSD
by Ornstein and F\"urth.
Setting $ G=0$ $(\Gamma=0)$ in our theory, we obtain the MSD for structureless particles of the 2-dimensional.
 It was a problem that they studied and derived the so-called Ornstein-F\"urth formula \cite{Ornstein1919,Langevin2017}, 
$$
 {\rm MSD}\equiv \langle\mbox{\boldmath $x$}^2(t)\rangle =4 D_{xx}
 \left[ t \mp {\displaystyle \frac{1}{\gamma}} ( 1-e^{-\gamma t} ) \right] , \eqno(20)
$$ where the upper sign $(-)$ is their result and the lower sign $(+)$ ours. 
A unique feature of their formula is that the MSD is $ \langle\mbox{\boldmath $v$}^2\rangle t^2 $
when $ t \ll 1/\gamma $ and $4 D_{xx}t$ when $ t \gg 1/\gamma $. Their explanation is that no scattering
takes place within a short period of time and therefore, the range is given by 
 $ \delta \mbox{\boldmath $x$}=\sqrt{\langle\mbox{\boldmath $v$}^2\rangle} \, \delta t$.
 Although the explanation is plausible, it is, in fact,  attributed to the choice of initial conditions
 that is not accessible by their method integrating the velocity autocorrelation function. As we have seen, the MSD satisfies the 2nd order differential equation
 that needs 2 initial conditions, the first $\langle\mbox{\boldmath $x$}^2(0)\rangle =0$ and the second 
 is $\langle\mbox{\boldmath $x$}^2(0)\rangle'=2Q_{\rm P}(0^+)$ given in the Eq.(16). As might be expected, we achieve the Ornstein-F\"urth formula
 if $q_{\rm P}(t)$ is the integrand instead of $Q_{\rm P}(t)$.  The formula has still remained a standard
 with which experimental data are analyzed \cite{Selmeczi2005,Selmeczi2014} and therefore, it is important to point out and give it a correction.

In conclusion, we developed a new formalism of Brownian motion and applied it to the problems of
Skyrmionic diffusion. A new type of diffusion coefficients
and corresponding diffusion processes were predicted. 
 In the formalism, the velocity and position distribution are treated on the same
footing, introducing the position-velocity correlation functions. 
We also pointed out the error in the famous Ornstein-F\"urth formula.

This research and development work was supported
by the Ministry of Internal Affairs and Communications.

\end{document}